\documentclass[preprint, times]{aastex62}
\usepackage{amsmath}
\usepackage{xspace}
\usepackage{graphicx}
\usepackage{epstopdf}
\usepackage{placeins}
\usepackage{lineno}

\newcommand{\msun}{\mbox{M$_\odot$}}
\newcommand{\degree}{\mbox{$^\circ$}}
\newcommand{\cxo}{\mbox{C$^{17}$O}}
\newcommand{\cxs}{\mbox{C$^{34}$S}}
\newcommand{\hcop}{\mbox{H$^{13}$CO$^{+}$}}
\newcommand{\chxoh}{\mbox{CH$_3$OH}}
\newcommand{\funit}{\mbox{mJy~beam$^{-1}$}}
\newcommand{\funitv}{\mbox{mJy~beam$^{-1}$~km~s$^{-1}$}}
\newcommand{\kms}{\mbox{km~s$^{-1}$}}

\shorttitle{EC~53}
\shortauthors{Lee et al.}

\begin{document}

\title{The circumstellar environment around the
    embedded protostar EC~53}

\author{Seokho Lee}
\author{Jeong-Eun Lee}
\affiliation{School of Space Research, Kyung Hee University, 1732 Deogyeong-daero, Giheung-gu, Yongin-si,Gyeonggi-do, Korea; jeongeunlee@khu.ac.kr}
\author{Yuri Aikawa}
\affiliation{Department of Astronomy, University of Tokyo, 7-3-1 Hongo, Bunkyo-ku, Tokyo 113-0033, Japan}

\author{Gregory Herczeg}
\affiliation{Kavli Institute for Astronomy and Astrophysics, Peking University, Yi He Yuan Lu 5, Haidian Qu, 100871, Beijing, PR China}

\author{Doug Johnstone}
\affiliation{NRC Herzberg Astronomy and Astrophysics, 5071 West Saanich Rd, Victoria, BC, V9E 2E7, Canada }
\affiliation{Department of Physics and Astronomy, University of Victoria, Victoria, BC, V8P 1A1, Canada}

\footnote{\bf{Seokho Lee moved to National Astronomical Observatory of Japan (seokho.lee@nao.ac.jp).}}

\begin{abstract}
EC53 is an embedded protostar with quasi-periodic emission in
the near-IR and sub-mm.  We use ALMA high-resolution observations of continuum and molecular
line emission to describe the circumstellar environment of EC
53.
The continuum image reveals a disk with a flux that suggests a mass of
0.075~\msun, much less than the estimated mass in the envelope,  and an in-band
spectral index that indicates grain growth to centimeter sizes. 
Molecular lines trace the outflow cavity walls, infalling and 
rotating envelope, and/or the Keplerian disk. The rotation profile
of the \cxo~3--2 line emission cannot isolate the Keplerian motion clearly 
although the  lower limit of  the protostellar mass can be calculated 
as 0.3~$\pm$~0.1~\msun\, if the Keplerian motion is adopted. 
The weak \chxoh\, emission, which is anti-correlated with the \hcop~4--3
line emission, indicates that the water snow line
is more extended than what expected 
from the current luminosity, attesting to bygone outburst events. The extended snow line may persist for longer at the disk surface because the
lower density increases the freeze-out timescale of methanol and water.
\end{abstract}


\section{Introduction}
 Low mass protostars form through the gravitational collapse of molecular 
cloud cores. The core material accretes onto protostars through a temporal 
reservoir, the accretion disk.  Historical star formation models with a 
constant accretion rate of a few 10$^{-6}$ \msun${\rm yr}^{-1}$
\citep[e.g.,][]{Shu1977,Masunaga2000} predict protostellar
luminosities that are much higher than the 
actual luminosities of most protostars 
\citep[e.g., ][]{Kenyon1990,Enoch2009,Dunham2010}. 
One promising solution of this luminosity problem is the episodic
accretion, although other time-dependent solutions are also possible 
(see the discussion of \citet{Hartmann2016} and \citet{Fischer2017}).

Several observational phenomena  support the episodic accretion directly 
or indirectly. Bursts in the optical brightness with various time scales 
are observed in FU Orionis objects \citep[e.g., ][]{Herbig1977,Herbig2008,Audard2014} 
and EXors \citep[e.g.,][]{Aspin2000,Kospal2008,Lorenzetti2012}.
 Knots along jets and clumpy structures in protostellar outflows 
\citep[e.g.,][]{Plunkett2015} could be also induced by accretion variability.
 In addition, the variation in accretion luminosity affects the chemistry 
in the disk and envelope \citep[e.g.,][]{Lee2007,Visser2015}; the 15.2~$\mu$m  
pure CO$_2$ ice absorption features toward low-luminosity protostars \citep{Kim2012} 
and the extended snow lines of volatiles compared to the theoretical snow line 
locations expected  from the current luminosity \citep{Jorgensen2013,Jorgensen2015} 
are explained by past bursts of accretion.  
Brightness variation in the sub-mm by the accretion burst is detectable
even for embedded protostars \citep{Johnstone2013}, as demonstrated for the Class 0 star HOPS 383 and two massive protostars \citep{Safron2015,Hunter2017,Liu2018}.

Although the mechanism of episodic accretion in protostars has not 
yet been securely identified, two main ideas are proposed to explain the 
large accretion bursts.  First, disk instabilities, such as thermal instability 
\citep{bell1994}, gravitational instability \citep[e.g., ][]{Vorobyov2010}, 
 and the combined effect of both magnetorotational instability and gravitational 
instability \citep{Zhu2009,Bae2014}, may trigger a large amount of mass to accrete
through the disk and onto the star.
 The other type of possible mechanism is the perturbation of the disk  by an 
external trigger, such as a binary companion \citep{Bonnell1992}, the tidal 
disruption of radially migrating gas clumps/giant planets \citep{Vorobyov2005,Nayakshin2012}, 
 or disk variability linked to a stellar magnetic cycle  
 \citep{Armitage1995,DAngelo2012,Armitage2016}.  
 The role of these  mechanisms may be more important at earlier stages in protostellar
growth, however any evolutionary change is difficult to evaluate with
most surveys, since the common optical and near-IR survey probe later stages in
the evolution of young stellar objects 
\citep[e.g.][]{Hillenbrand2015,Contreras2017,Contreras2019}.

Recently, the JCMT-Transient monitoring survey has been searching for
variability at the protostellar stage with monthly monitoring of
sub-mm continuum emission in eight nearby star-forming regions  \citep{Herczeg2017}.    
 The first sub-mm variable detected in this survey, EC~53 \citep{Yoo2017}, 
stands out in amplitude relative to other protostars in the 8 fields 
 \citep{Mairs2017, Johnstone2018}.
EC~53, first identified by \citet{Eiroa92} as a member of the Serpens
Main Cloud \citep[distance of $436\pm9$ pc][]{Ortiz2017}, is an embedded 
protostar with a bolometric luminosity of 1.7 -- 4.8 L$_\odot$ 
\citep{Evans2009,Enoch2009,Gutermuth2009,Dianatos2010,Dunham2015}.
Past near-IR monitoring shows that the K magnitude of EC~53 varies  with 
a period of $\sim$543 days \citep{Hodapp1999,Hodapp2012}.
 The sub-mm observations  \citep{Yoo2017} confirm this periodicity 
and strongly suggests that this sub-mm variability should be induced by 
the variable accretion rate. In addition, this short $\sim$1.5 years 
variability may be related with a very close companion or a protoplanet, 
which can induce the periodic gravitational instability in the disk \citep{Lodato2004}.  
This short-period brightness variation of an embedded protostar in sub-mm 
casts an important question on when planets start to form.

The range of instabilities and the consequences of outbursts are both
expected to leave traces in the circumstellar environment in the
immediate vicinity of the protostar.  To better describe the
kinematics and chemical properties of this environment, we have
obtained high-resolution Atacama Large Millimeter/submillimeter Array
(ALMA) imaging of gas and dust.  

\section{Observations}
 The observation of EC~53 was carried out with ALMA during Cycle 4 
(2016.1.01304.T, PI: Jeong-Eun Lee) on 2017 May 21. 
 The four spectral windows were set for 336.995 -- 337.229 GHz, 
338.345 -- 338.579 GHz, 346.945 -- 347.180 GHz, and 349.396 -- 349.630 GHz, 
each with a spectral resolution of 244.141 kHz ($\sim$0.2 \kms). 
This setup contains \cxo\, 3--2, \chxoh\, 7$_{\rm k}-6_{\rm k}$, \hcop\, 4--3, 
CCH N=~4--3, J=~7/2--5/2, F=~4--3 and F=~3--2 lines.  
 Forty-one 12-m antennas were used in the C40-5 configuration, with 
baselines in the range from 15.1 m to 1.1 km. The total observing 
time on EC~53  was 48.1 minutes.

 The data were initially calibrated using the CASA 4.7.2 pipeline \citep{McMullin2007}.
 The quasar J1751+0939 was used as a bandpass and amplitude calibrator, 
and the nearby quasar J1830+0619  was used as a phase calibrator.
Self-calibration was applied for better imaging.

 For  molecular lines, natural weighting with a UV-taper of 0.2\arcsec\, 
was used to obtain images with a higher signal-to-noise ratio and a 
resolution of 0.3\arcsec. 
A spectral resolution of 0.25 \kms\, was used for {\it strong} \cxo, 
\hcop, and CCH lines, while 0.5 \kms\, was used for {\it weak} lines 
to reduce the RMS noise (Table~\ref{tab:mol_lines}).  
The RMS noise level for the strong molecular lines is  $\sim$4 \funit\, near 
the continuum peak. 
 The images of \cxo, \hcop, and two strong CCH lines were produced by Briggs 
weighting with a robust factor of 0.5 to obtain higher resolution (0.2\arcsec).

 A dust continuum image was produced  using line-free channels and was cleaned with 
two weightings: (1) natural weighting with the UV-taper of 0.2\arcsec\, to 
compare with the molecular line images and (2) uniform weighting  only with 
the UV distances longer than 500 k$\lambda$ to obtain a higher spatial resolution, 
with a beam size of 0.15\arcsec\, $\times$ 0.11\arcsec\, oriented at position angle of  -88.4\degree.
The RMS noise levels  of the two images are 0.21 and 0.38 \funit, respectively.  
The data are analyzed in the image plane.

\section{Results}
\subsection{Dust continuum images} \label{sec:dust}
\begin{figure*}
\includegraphics[width=1.00 \textwidth]{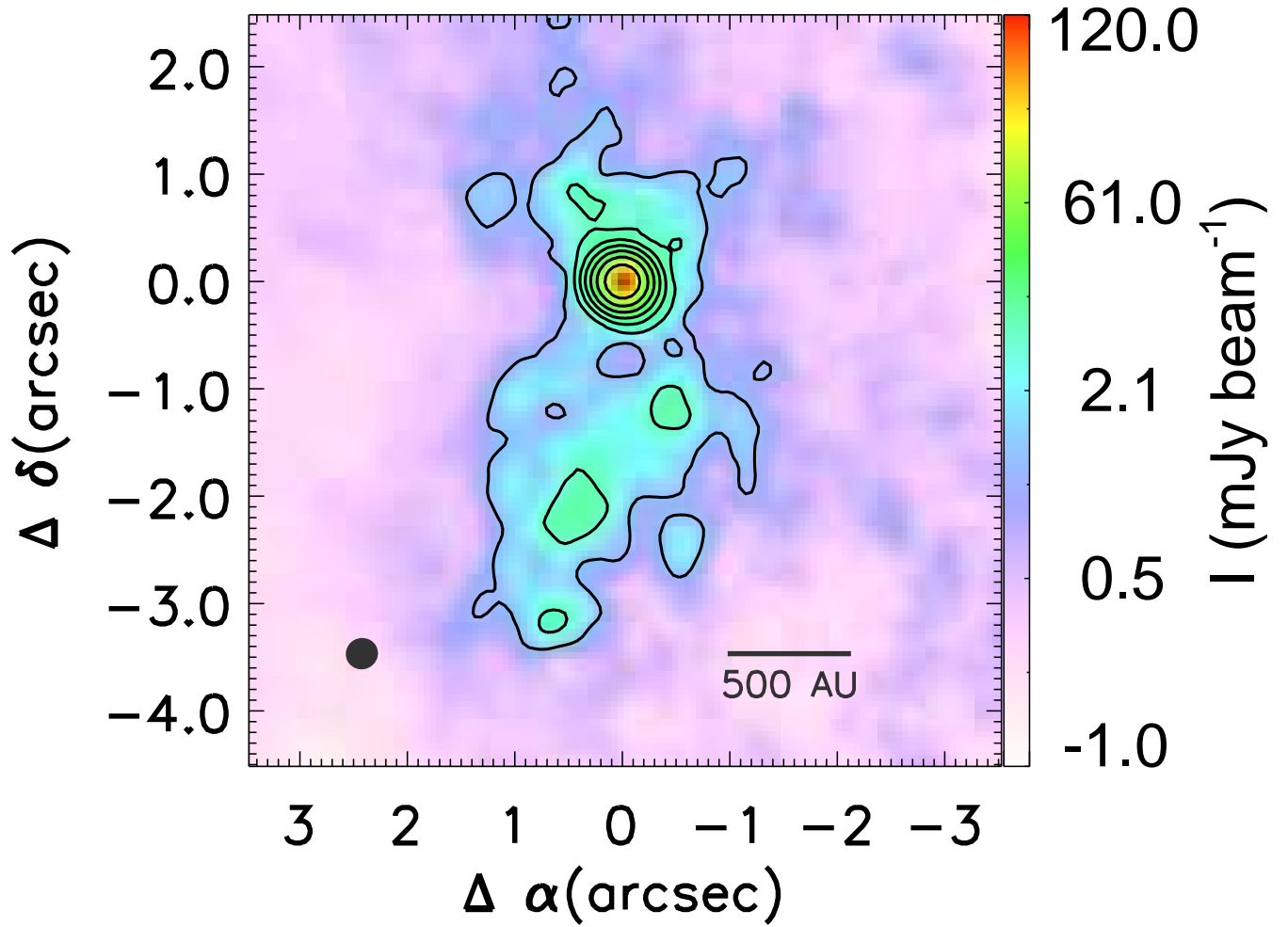}
\caption{The 873 $\mu$m dust continuum image produced with the natural weighting 
  and the UV taper of 0.2\arcsec. 
  The contour levels increase by a factor of two from 5 $\sigma$ to 320 $\sigma$ 
 (1 $\sigma$= 0.21 \funit).
  The origin is the position of the continuum peak for EC~53. 
  The ellipse on the bottom left corner represents the synthesized beam, 
   and the scale bar of 500 AU is presented at the bottom right. 
  Note that the color scale has two linear slopes below and above 2.1 \funit\, in order to
	present the extended emission.
}
\label{fig:cont_natural}
\end{figure*}

\begin{figure*}
\centering
\includegraphics[width=0.45 \textwidth]{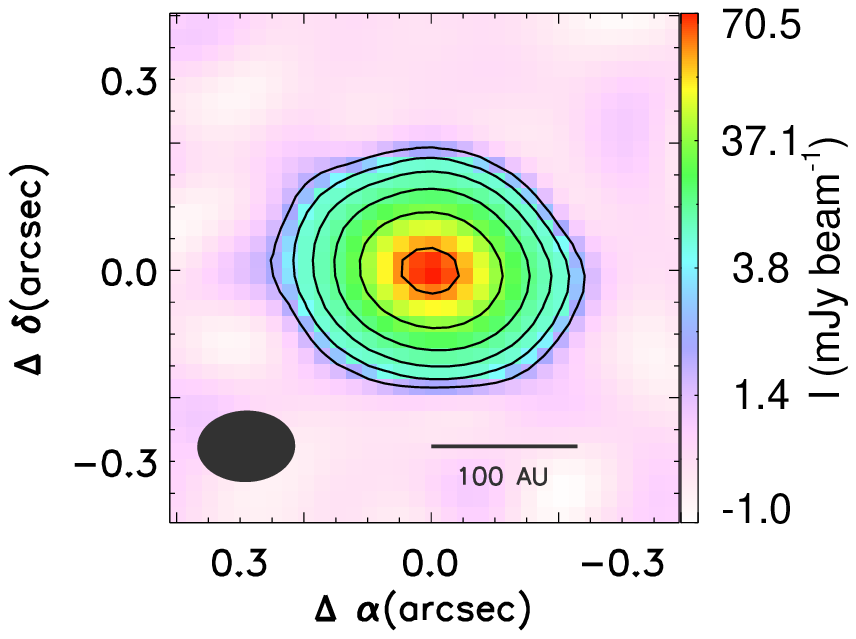}
\includegraphics[width=0.45 \textwidth]{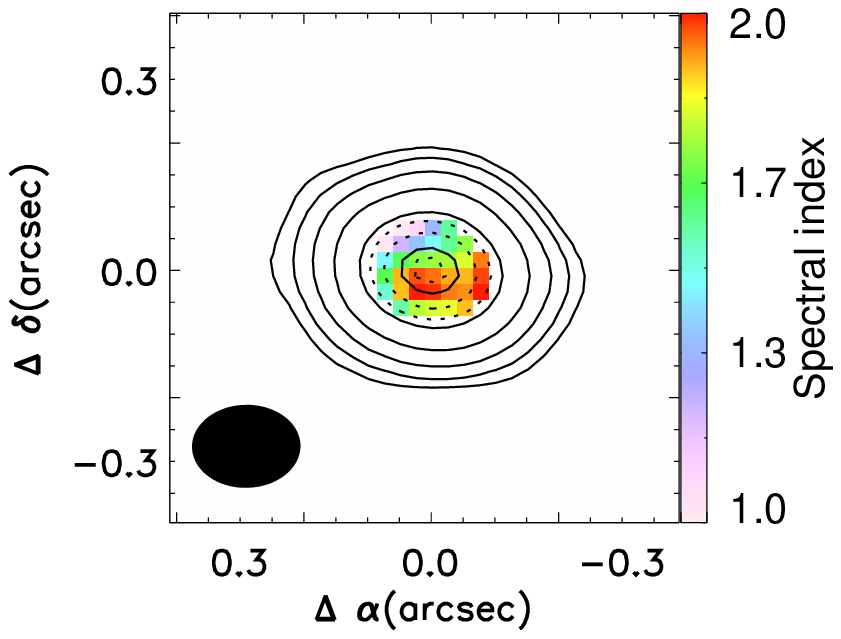}
\caption{ Left: Dust continuum images produced with the uniform weighting.
 The contour levels increase by a factor of two from 5 $\sigma$ to 160  $\sigma$ 
  (1 $\sigma$= 0.38 \funit). 
  Note that the color scale has two linear slopes below and above 3.8 \funit.  
	The black ellipse on the bottom left corner indicates the beam size, 
	and the scale bar of 100 AU is presented at the bottom right.  
 Right: The distribution of the spectral index ($\alpha$; F$_\nu\propto \nu^\alpha$)
  derived from the images produced with the uniform weighting. The solid contours 
  are the same as those on the left panel. 
  The dotted contours represent the 1~$\sigma$ errors of the spectral index from 0.5 (outermost) 
  to 0.3 (innermost) with the step size of 0.1. 
   The beam size of the image for the spectral index is  presented as
  the black ellipse on the  bottom left  corner.
}
\label{fig:cont_uniform_alpha}
\end{figure*}

The 343.4 GHz (873 $\mu$m) continuum image with a lower resolution 
(Figure~\ref{fig:cont_natural}) shows extended emission along the north-south 
direction as well as a compact source. The compact source is resolved in the higher 
resolution image as shown on the left panel of Figure~\ref{fig:cont_uniform_alpha}.
The compact emission could originate from the disk (see  
Sections~\ref{sec:model} and \ref{subsec:dust_evol}). 
The properties of the continuum emission are derived by fitting a Gaussian profile 
using the task `imfit' in CASA. 

The peak of the continuum emission is located at 
$\alpha$(J2000)= 18$^h$29$^m$51.17685$^s$ 
and $\delta$(J2000)= +01$^d$16$^m$40.3892\arcsec.
This peak position is consistent to within $\sim$0.2\arcsec\ of the positions from {\it Spitzer} \citep{Harvey2006,Hsieh2013}, VLA \citep{Ortiz2015}, and high-resolution Keck $K_S$-band imaging \citep{Hodapp2012}, after applying the proper motion correction for Serpens Main of $\sim$10~mas yr$^{-1}$ towards the south-east \citep[]{Ortiz2017,Herczeg2019}. The position is offset to the north of the 2MASS \citep{Cutri2003} and neoWISE \citep{Cutri2014} positions, likely a result of nebulosity seen to the south of the EC~53 binary system by \citet{Hodapp2012}.

The deconvolved image of the disk shows that the major and minor axes are 
144.2 $\pm$ 2.4 mas and 118.4 $\pm$ 2.2 mas with a position angle of  
60.0 $\pm$ 4.2\degree. 
The ratio of the major and minor axes suggests the disk inclination of 
34.8 $\pm$ 2.1\degree. When this position angle and the inclination are adopted, 
the disk radius that encircles 95\% of the disk flux is 214 mas 
(93 AU at the distance of 436 pc), similar to the typical size of more
evolved protoplanetary disks \citep[e.g.][]{Tripathi2017,Cieza2019,Long2019}.

This position angle and inclination are both consistent with previous observations.
 The observed position angle of the disk is roughly perpendicular to the
direction of the H$_2$ jet at  140\degree\ \citep{Herbst1997} 
and  the $^{12}$CO 3--2 outflow at 135\degree\ \citep{Dianatos2010}.
\citet{Hodapp2012} suggested that the disk of EC~53 is seen nearly edge-on 
based on the cometary feature in the near-IR image. 
 However, the inclination of the outflow axis could be lower than the 
half-opening angle of the outflow cavity when a protostar is observed in 
near-IR scattered light \citep{Reipurth2000}. 
Indeed, the near-IR image of EC~53  \citep[Figure 6 in][ see 
the blue dashed curve on the bottom right panel in Figure~\ref{fig:mol_mom0}]{Hodapp2012} 
indicates that the half-opening angle of the outflow cavity is smaller than 45\degree\, 
 and the inclination is between 30\degree\,and 45\degree\, 
 \citep[see Figure 17 in ][]{Reipurth2000}.
The best-fit spectral energy distribution model of EC~53 also {\bf requires} a small inclination of 30\degree\, 
 (Baek et al. submitted). 
 Furthermore, for Class I sources, [Fe II] forbidden lines trace high velocity gas 
with a radial velocity of 100 -- 200 \kms\, \citep{Davis2011}.  
Near-IR spectra of EC~53 show that the  1.644~$\mu$m [Fe II] line is blue-shifted 
by $\sim$170 \kms\, (S. Park et al. in prep), which implies that the inclination 
of  the jet should be close to face-on.

 The peak intensity and total flux of the disk, which are measuring by applying `imfit' 
to the left image of Figure~\ref{fig:cont_uniform_alpha}, 
are 69.3 $\pm$ 0.4 \funit\, and 139.7 $\pm$ 1.2 mJy, respectively.
 The compact emission shown in the lower resolution image (Figure~\ref{fig:cont_natural}) 
has the flux of 155 mJy within an aperture of 0.7\arcsec, 
implying that the compact emission is mostly radiated from the disk shown in 
the high resolution image (Figure~\ref{fig:cont_uniform_alpha}).  
On the other hand, the total continuum flux in Figure~\ref{fig:cont_natural} is 330~mJy.
 The 850 $\mu$m flux (in the beam of 14.6\arcsec) of EC~53 with the JCMT/SCUBA-2 is 
960 mJy at the quiescent phase \citep{Yoo2017}, and thus, at least 60\% of the continuum
emission arises from an extended envelope and is resolved out in this ALMA observation. 

 The disk mass may be estimated from the continuum flux by
\begin{equation} \label{eq:one}
   M_{\rm disk} = \frac{F_\nu d^2}{\kappa_\nu B_\nu (T_{\rm dust})},
\end{equation}
where $F_\nu$ is the observed flux (139.7 $\pm$ 1.2 mJy), $d$ is the distance to EC~53, 
$\kappa_\nu$ is the dust opacity, and $B_\nu (T_{\rm dust})$ is the Planck function 
with the dust temperature of $T_{\rm dust}$.  Adopting an average dust
temperature of 20 K \citep[following][]{Andrews2005},
a dust opacity 
$\kappa_\nu =$ 3.4 cm$^{2}$~g$^{-1}$ \citep{Beckwith1990}, and a
gas-to-dust ratio of 100 leads to a disk mass of 0.075 $\pm$ 0.003
M$_\odot$.  This estimate for disk mass is consistent with standard
techniques but depends on highly uncertain assumptions, including the
dust opacities and optical depths \citep[see review by][]{Williams2011}.

 The right panel of Figure~\ref{fig:cont_uniform_alpha} shows the spectral index 
$\alpha$ between two continuum images produced with the uniform weighting at 
frequencies of $\nu_1$ = 348.38 GHz and  $\nu_2$ = 337.86 GHz.  
 The spectral index has a relation with the optical depth $\tau_{\rm 343.4 GHz}$ 
and the power-law index of the dust opacity $\beta$ as
 \begin{eqnarray}
  \frac{F_{\nu_1}}{F_{\nu_2}}&=& \left(\frac{\nu_1}{\nu_2}\right)^\alpha 
  = \frac{B_{\nu_1}(T)(1-\exp(-\tau_{\nu_1}))}{B_{\nu_2}(T)(1-\exp(-\tau_{\nu_2}))},
  \end{eqnarray}
 \begin{equation}
   \tau_{\nu_i}= \tau_{\rm 343.4 GHz } \left(\frac{\nu_i}{\rm 343.4 GHz}\right)^\beta,
 \end{equation}
 where $F_{\nu_i}$ and $\tau_{\nu_i}$ are the flux and optical depth at the frequency $\nu_i$.
The spectral index at the continuum peak is 1.9 $\pm$ 0.3, which implies that either 
the disk is optically thick at 343.4 GHz and/or that dust grains have grown 
(see Section~\ref{subsec:dust_evol}).

The JCMT/SCUBA-2 850 $\mu$m flux of EC~53  after removing the contribution of 
the disk implies the envelope mass is 0.86 -- 1.25 \msun\, 
\citep[see Equation (1) in ][]{Jorgensen2009}. 
The disk to envelope mass ratio, 6 -- 9\%, implies that EC~53 could be a late 
Class 0 (1 - 10\%) or an early Class I source (20 -- 60 \%) \citep{Jorgensen2009}.
 This is consistent with the early Class I stage \citep{Dunham2015} derived from the 
bolometric temperature  ($T_{\rm bol}$= 130~K) of the SED, although a continuum radiative transfer 
model is needed to estimate more accurate disk and envelope masses.

\subsection{Molecular lines}

In this section, we infer the morphology of the circumstellar material by measuring gas kinematics in the spectral images 
of emission lines. The different emission lines probe the disk, infall, and outflows of the source, as described below.
A source velocity of 8.5~\kms\, is 
adopted throughout this analysis, based on the rotation profile 
of the \cxo\, line along the disk direction (Section~\ref{sec:model}). 
 This value is similar to a previous measurement of $\sim$8.35~\kms\, 
using N$_2$H$^+$ 1--0 line emission measured with the CARMA 
(Combined Array for Research in Millimeter-wave Astronomy) data that 
has a beam size of 7\arcsec\, \citep{Lee2014}, with a slight offset
that may be explained because the N$_2$H$^+$ species is depleted
at the source position (Figure~\ref{fig:classy_n2hp}).

\begin{figure*}
\includegraphics[width=1.00 \textwidth]{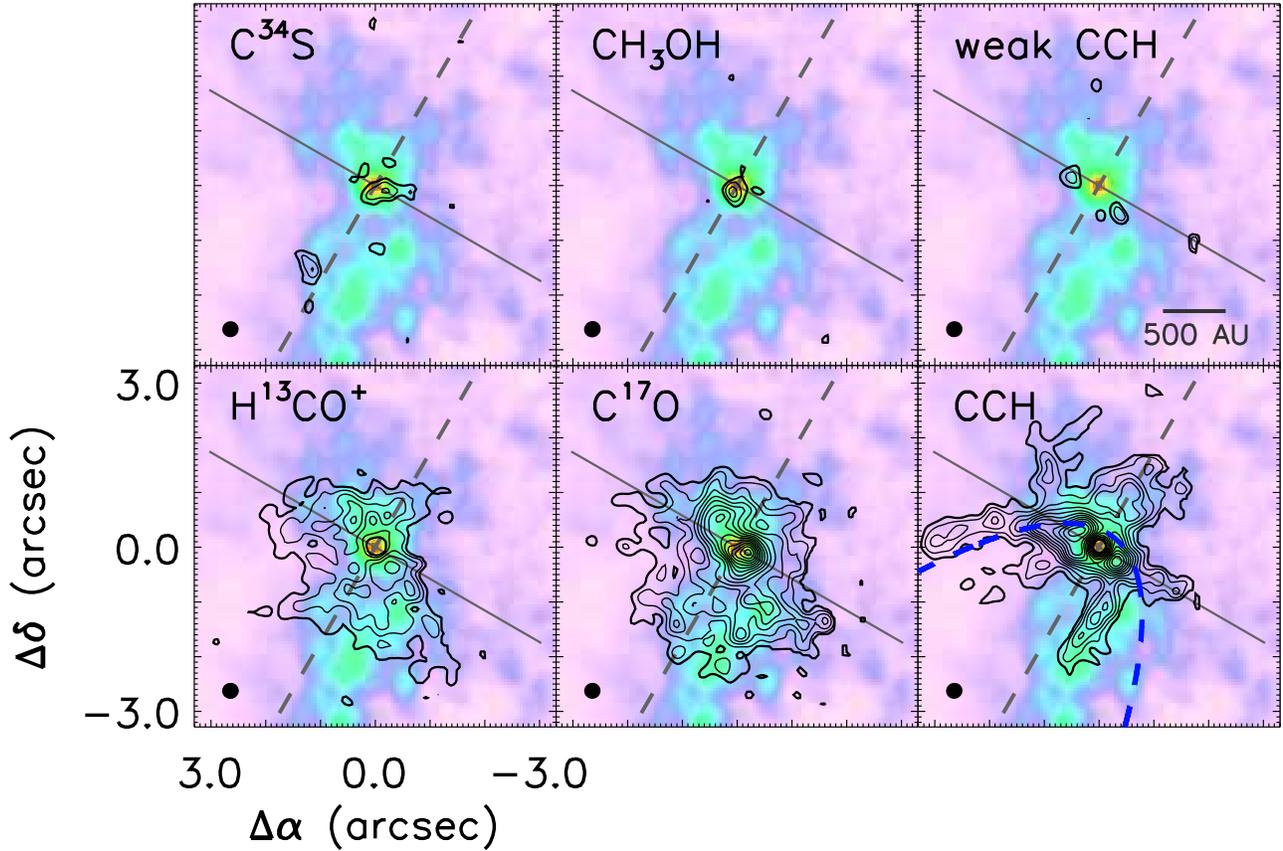}
\caption{Integrated intensity maps for the detected molecular lines toward EC~53. 
 The background image is the zoom-in of Figure~\ref{fig:cont_natural}. 
 The thick solid and dashed  gray lines indicate the disk and outflow directions, respectively.
 The blue dashed curvature in the bottom right  panel indicates the outflow cavity wall 
traced by a near-IR observation \citep{Hodapp2012}.
 The ellipses on the bottom left corner represent the synthesized beam.
The scale bar of 500 AU is at the bottom right of the weak CCH map.
 The \chxoh\, and weak CCH lines are marginally detected (see Table~\ref{tab:mol_lines}). 
 Those lines of the individual molecules are stacked in order to get a higher 
 signal-to-noise ratio. 
  The 1 $\sigma$ RMS noise, the lowest contour, and subsequent contour step for 
  the individual molecules are as follow:
    4.7~\funitv, 3 $\sigma$, and 1 $\sigma$ for \cxs, 
    2.7~\funitv, 3 $\sigma$, and 1 $\sigma$ for \chxoh, 
    4.4~\funitv, 3 $\sigma$, and 1 $\sigma$ for weak CCH, 
    4.6~\funitv, 5 $\sigma$, and 2 $\sigma$ for \hcop,
    4.3~\funitv, 5 $\sigma$, and 3 $\sigma$ for \cxo,
    5.6~\funitv, 5 $\sigma$, and 3 $\sigma$ for CCH.
}
\label{fig:mol_mom0}
\end{figure*}

 Figure~\ref{fig:mol_mom0} shows integrated intensity maps for the 
detected molecular lines toward EC~53. \hcop\, 4--3, \cxo\, 3--2, and 
CCH N= 4--3, J= 7/2--5/2, F= 4--3, F= 3--2 (in the bottom panels) 
consist of compact emission near the continuum peak and more extended 
emission while C$^{34}$S 8--7, \chxoh, and other weak CCH lines 
(in the top panels) show  only compact and weak emission 
near the continuum source.  
 Those weak \chxoh\, and CCH lines are stacked individually to get 
a higher signal-to-noise ratio. The observed lines are listed in 
Table~\ref{tab:mol_lines}.

 The integrated intensity maps of the \hcop\, and \cxo\, lines generally 
show similar distributions.
 Both lines have red-shifted and blue-shifted components in the north and south, 
respectively, as shown in the intensity weighted velocity map (Figure~\ref{fig:mol_mom1}).
 The velocity channel maps of \cxo\, and \hcop\, (Figures~\ref{fig:chan_c17o} 
and \ref{fig:chan_hcop}) show that the red-shifted components (9.00 -- 9.75 \kms) 
have decreasing velocity with the distance from the protostar.  Such a profile is expected for 
 red-shifted emission from infalling (accreting) 
material, as also seen in the Class I source L1489 IRS \citep{Yen2014}. 
 The same trend is also observed in the \cxo\, blue-shifted components 
(8.00 -- 7.25 \kms) near the protostar (r $<$ 1\arcsec).
On the other hand, the  blue-shifted velocity (8.50 -- 7.75 \kms) increases 
with the distance from the protostar at r $\geq$ 1\arcsec. 
 This implies that the blue-shifted emission at the scale of r $\geq$ 1\arcsec\,  
traces the gas in the outflow cavity wall \citep[e.g.,][]{Arce2013}. 
 The blue-shifted monopolar outflow could be due to an asymmetric structure of 
the circumstellar environment \citep{Loinard2013}. 

 At the continuum peak, absorption against the continuum emission is detected in 
the \hcop\, and strong CCH lines, while self-absorption is detected in the \cxo\, 
line (Figure~\ref{fig:mol_spec_cen}). 
This deep absorption below the continuum level in the \hcop\, and 
strong CCH lines is red-shifted, tracing the infalling envelope material \citep{evans2015}.
  The high velocity ($|\Delta V| > 1$ \kms\, from the source velocity)  
emission feature detected in the \cxo\, line (the red spectrum in 
Figure~\ref{fig:mol_spec_cen}),  which may trace the disk rotation, 
is missing in the \hcop\, line. 
  Figure~\ref{fig:chan_hcop} shows that there is no \hcop\, line emission 
within $\sim$0.2\arcsec\, from the protostar.
  The related chemistry for this phenomenon is discussed in  Section~\ref{subsec:watersnow}.

 The strong CCH line emission (the bottom right contour image of 
Figure~\ref{fig:mol_mom0}) seems to trace the outflow cavity walls.  
 The stream to the east, which appears distributed along the blue-shifted 
outflow cavity wall, is also seen in the near-IR narrow band image covering the H$_2$ emission \citep{Hodapp2012}.
 Another stream toward the north direction looks like a mirror image 
of the eastward stream with respect to the disk midplane (the solid 
gray line in Figure~\ref{fig:mol_mom0}), which could  imply that the 
northward stream might trace the red-shifted outflow cavity wall.  
CCH emission along the outflow cavity walls is also found in L1527, IRAS 
15398-3359, and L483 \citep{Oya2014,Sakai2014,Oya2017}.  
 
 The \chxoh\, and other weak CCH lines are marginally detected 
(see Table~\ref{tab:mol_lines}). 
 While the \chxoh\, lines are detected near the continuum peak, 
the weak CCH lines are missing toward the continuum source, 
instead peaking at 0.55\arcsec\, from the continuum peak  along 
the disk direction.
 The same phenomenon has been observed in other sources: in  L1527,
  IRAS 15398-3359, and L483 the CCH is also depleted near 
the continuum peak and has strong emission off the peak
\citep{Oya2014,Sakai2014,Oya2017}.  
 The CCH molecule reacts with H$_2$ at warm temperatures 
\citep[see e.g.,][]{Aikawa2012} and could be destroyed by those 
gas phase reaction  in the warm and dense inner envelope and/or 
frozen on the dust grain in the cold and dense disk. 
  
 The C$^{34}$S 8--7 line seems to trace the dense molecular gas 
near the continuum peak  as well as in the outflow cavity wall. 
 The SiO 7--6 line was also observed but not detected.

 \subsubsection{Kinematics of \cxo\, 3--2: Disk vs. Infalling rotating envelope?} \label{sec:model}
 The rotation profile along the disk direction provides evidence of the 
kinematics of the disk. 
Figure~\ref{fig:chan_c17o_zi} shows velocity channel maps for \cxo\, 
near the continuum peak.
 The innermost regions have a position-velocity distribution that
 would be expected for a disk.  Beyond 0.5\arcsec\, along the position 
angle of 60\degree\, (the black solid line in  Figure~\ref{fig:chan_c17o_zi}),  
the spatial distribution of red-shifted components (9.75 and 9.50 \kms) 
 implies that this emission might not be associated with the disk. 
 In the following  analysis, we fit the rotation profile 
only within 0.5\arcsec. The highest velocity components are close to the 
protostar, and the models with different position angles from 30\degree\, 
to 60\degree\, do not show significant variation in the rotation profile.

 Figure~\ref{fig:c17o_pv} shows the position-velocity diagram along 
the disk (left) and the rotation velocity profiles (right) of EC~53. 
 The power-law index that fits the velocity profile is  -2.1 $\pm$ 2.2.
 If the highest blue-shifted point is  excluded, the best-fit power
 law is  -1.1 $\pm$ 0.7, indicative of the rotating infalling envelope  
 under the constraint of angular momentum conservation \citep{Sakai2014}.
  However, our current observation  with large scatter and  a low 
signal-to-noise ratio cannot rule out  a rotational profile produced by 
the Keplerian disk.
 The range of the rotation velocity corresponds to  the Keplerian motions 
 with the protostellar masses of 0.2 and 0.4 \msun\, when we adopted the 
 inclination of 35\degree.
 Since the protostar is still young and the envelope massive, the present-day protostellar mass may be much lower than the final mass after stellar assembly is complete.
   
\section{Discussion}
\subsection{Dust grain growth}\label{subsec:dust_evol}
 The observed 873 $\mu$m continuum emission is mainly from dust thermal emission.
 In the continuum images produced with uniform weighting, the peak intensity and 
the spectral index at the continuum peak are 69.3~$\pm$~0.4 \funit\, 
and 1.9~$\pm$~0.3, respectively. 
The spectral index lower than 2 cannot be explained by non-thermal emission, 
since VLA 4.5 and 7.5 GHz observations 
\citep[with a beam size of $\sim$0.4\arcsec,][]{Ortiz2015}
indicate that the 873 $\mu$m flux from the non-thermal emission is lower than 0.2 mJy.
 Therefore, the spectral index of 1.9~$\pm$~0.3 implies that the continuum emission is optically thick
 and/or that the power-law index of dust opacity is lower than 1 in the isothermal condition.

 The sub-mm continuum spectral index can be used to derive the  properties 
 of emitting sources and dust grains.
   The low spectral index (1.9~$\pm$~0.3) for the dust continuum emission
indicates that the compact continuum emission could be emitted from the disk.
 On the other hand, the spectral index derived from the ACA (Atacama Compact 
Array) images of EC~53 in band 6 and 7, which trace the envelope of EC~53, 
 is about 2.5 (W. Park, in prep.).
Continuum emission images at 1.1 mm and 3 mm toward two Class I sources Elias~29 
and WL~12 show that the spectral indices in the disk are typically lower than 2,
while in the envelope are typically higher than 2.5 \citep{Miotello2014}. 
Figure~\ref{fig:tau_beta} shows the optical depth and the power-law index of 
dust opacity, $\beta$, for given dust temperatures reproducing the observed 
spectral index of 1.9~$\pm$~0.3. When we adopt the dust temperature of  20 K and 
$\beta$= 1, which are used for the calculation of the disk mass, the average 
optical depth of dust continuum is 1.8$_{-1.0}^{+2.2}$ 
(see the blue circle in Figure~\ref{fig:tau_beta}). 
If this optical depth is considered, the disk mass derived by Equation~(\ref{eq:one}) 
is smaller than the actual value by a factor of two.
On the other hand, if we know the dust temperature and the optical depth 
of dust continuum, then we can derive $\beta$, which constrains the dust 
properties; grain growth in the disk reduces the power-law index of dust 
opacity \citep[e.g.,][]{Natta2007}.

 The optical depth of the dust continuum  can be derived from a molecular line.
 As shown in the integrated intensity maps for the detected molecular lines 
(Figure~\ref{fig:mol_mom0}), the emission peaks of all the molecular lines
are off the continuum peak. 
 Figure~\ref{fig:c17o_uniform} shows  the  \cxo\, 3--2 intensity map 
integrated over  the velocity from 1.0 to 1.5~\kms\,  with respect to 
the source velocity.
 This integrated intensity map was produced using the uniform weighting to 
trace how the \cxo\, emission is distributed  near the continuum peak, by 
excluding  the contribution of the  extended emission.
 The \cxo\, line emission is clearly depleted near the continuum peak and 
peaks around 0.13\arcsec\, from the continuum peak.
 This phenomenon was also observed in TMC~1A \citep{Harsono2018} and V883~Ori 
 \citep{Lee2019}.
 A high optical depth and strong emission of the dust continuum can reduce 
the intensity of the molecular line emission;  the observed line intensity 
is reduced by $\exp(-\tau_c)$  with the dust optical depth of $\tau_c$ 
\citep{Yen2016,Lee2019}. 

 We adopt a simple thin disk with power-law distributions for the continuum 
optical depth $\tau_c(r)$ and the dust temperature $T_{\rm dust}(r)$:
 \begin{eqnarray}
   \label{eq:five}
   \tau_c(r) &=& \tau_{\rm 0.1} \left(\frac{r}{0.1\arcsec}\right)^{-\psi}, \\
   T_{\rm dust}(r)& =& T_{\rm 0.1} \sqrt{\frac{0.1\arcsec}{r}},
  \end{eqnarray}
  where $\tau_{\rm 0.1}$ and $T_{\rm 0.1}$ are the continuum optical depth 
and dust temperature at 0.1\arcsec. We test  three power-law indices of optical 
depth: $\psi$= 1.0, 1.5, and 2.0.
 The solid lines in Figure~\ref{fig:disk_model} indicate the models 
producing the observed 873~$\mu$m intensity at the continuum peak 
(69.3 $\pm$ 0.4 \funit).
 The emission from the inclined thin disk (inclination=~35\degree) are 
 convolved with the observed beam.

 The \cxo\, intensity within 0.5\arcsec\, (Figure~\ref{fig:chan_c17o_zi}) 
is lower than 45 \funit, which corresponds to a brightness temperature ($T_{\rm b}$) 
of $\sim 14$~K. If the  gas and dust have the same temperature, the \cxo\, brightness temperature is about two times lower than the gas temperature, indicating that the \cxo\, line is optically thin. 
 When we assume that the column density ratio of \cxo\, and dust is 
spatially constant, and the \cxo\, molecules are in the local thermal 
equilibrium (LTE) condition, the above thin disk model produces the \cxo\, 
emission distribution $I(r)$:
  \begin{equation}
   I(r) \propto \tau_c(r) \cdot n_u \left(T_{\rm dust} (r) \right) \cdot \exp\left(-\tau_c(r)\right),
     \label{eq:six}
  \end{equation}
 where the upper level population $n_u(r)$ is derived from the Boltzmann 
distribution with $T_{\rm dust} (r)$. 
 In the right hand side of Equation~(\ref{eq:six}), the first term 
($\tau_c(r)$) represents the total column density of \cxo, the second 
term ($n_u\left(T_{\rm dust}(r)\right)$) is the fraction of \cxo\, to 
produce the \cxo\, 3--2 line, and the third term indicates the extinction 
of the line by dust grains.

Our models with the dust temperature of  50 -- 55~K and dust optical 
depth  of 1.1 -- 1.6 at the radius of 0.1\arcsec\, can reproduce the 
\cxo\, emission peak around  0.13\arcsec\, (see the circles in 
Figure~\ref{fig:disk_model} and Table~\ref{tab:disk}). 
 The thin dust disk with a truncated radius of 0.13\arcsec\, can 
reproduce the observed continuum flux of 139.7~mJy as well as the 
disk size measured with `imfit'. 
 The entire dust disk is optically thick, and  thus, the \cxo\, emission
peaks at the dust disk radius. When the dust opacity of 
3.4 cm$^{2}$~g$^{-1}$ at 343.4~GHz \citep{Beckwith1990} and gas-to-dust 
ratio of 100 are adopted, the models show the disk mass is 0.1~\msun\, 
 with an uncertainty by a factor of two, which is similar to the disk 
mass of 0.075~\msun\ (see Section~\ref{sec:dust}). 
 Note that grain growth  can change the dust opacity \citep[e.g.,][]{Ricci2010}, 
and thus, the disk mass. The calculated dust mass would also be 3.3 times lower if we use $50$ K instead of $20$ K for the average dust temperature in the disk.

 For the dust temperature of $\sim$50~K and the continuum optical depth 
of $\sim$1.4 near the continuum peak in our model, the power-law index 
of the dust opacity ($\beta$) must be lower than 0.2~$\pm$~0.6 to reproduce the 
observed spectral index of 1.9~$\pm$~0.3 (see the black circle in Figure~\ref{fig:tau_beta}).
 This suggests that the maximum size of dust  grains should be larger 
than centimeter sizes \citep[e.g.,][]{Ricci2010,Testi2014}. 
 The  large disk mass of about 0.1 \msun\, and the large grain size imply 
a possibility of planet formation even in the early evolutionary stage, as 
argued by \citet{Harsono2018}.

\subsection{Water and CO snowlines} \label{subsec:watersnow}
Figure~\ref{fig:spec_map} shows the spectral maps of \hcop\, and \cxo\, 
on top of the image of the  \chxoh\, integrated intensity map.  
 Outside of the \chxoh\, emission area, the \hcop\, line has a similar 
intensity to the \cxo\, line. However, the \hcop\, line emission is 
depleted near the continuum peak, where the methanol lines emit. 
On the other hand, the \cxo\, line is still strong and broad near 
the continuum peak where the \hcop\, emission completely disappears. 
The \chxoh\, line might emit from the  disk or infalling rotating inner 
envelope  traced by \cxo\, because the \chxoh\, line has a similar line 
width to that of the \cxo\, line. 
The emission peak of \chxoh\, is offset from the continuum peak probably
 due to the optically thick continuum emission, as also shown in the \cxo\, emission distribution.
The anti-correlation between the \hcop\, and \chxoh\, emission was also 
detected in IRAS 15398-3359 and may be caused by the destruction of 
\hcop\, by water, which evaporates simultaneously with \chxoh\, 
\citep{Jorgensen2013}. The methanol emission is  known to be related 
closely with water emission \citep{Bjerkeli2016b}.
However, the observed \chxoh\, emission is extended up to $\sim$0.35\arcsec\, 
(see Figure~\ref{fig:spec_map}), which is an order of magnitude larger 
than the  water snow line expected from the current luminosity. 
In the thin disk model described above, the water snow line (100~K) 
is located around 0.03\arcsec.

 To measure the abundance of methanol, we first extracted the  \chxoh\, 
spectra with an aperture of 0.6\arcsec.  We then reproduced this spectrum with a simple LTE model 
using molecular data from the Cologne Database for Molecular Spectroscopy 
 \citep[CDMS,][]{CDMS2001}.  The model includes the dust continuum and opacity and assumes a line width of 2 \kms.
 The molecular hydrogen column density of $\sim$10$^{25}$ cm$^{-2}$  
is estimated from the dust continuum flux measured with the same 
0.6\arcsec\, aperture when the dust opacity of 3.4 cm$^{2}$~g$^{-1}$ 
\citep{Beckwith1990}, gas-to-dust ratio of 100, and the dust temperatures 
of 30, 50, and 70 K are adopted.
 The  derived methanol column density and abundance are $\sim$10$^{15}$ cm$^{-2}$ 
and $\sim$10$^{-10}$, respectively (see Figure~\ref{fig:spec_ch3oh}). 

The methanol emission from the much larger outburst of V883 Ori provides a useful comparison for that from EC 53.
The derived methanol abundance for EC 53 is two orders of magnitude lower than the \chxoh\, 
abundance of $\sim$10$^{-8}$ for V883 Ori \citep{Lee2019} although the physical size of the methanol emission regions are similar for V883~Ori and EC~53.
In both cases, the methanol abundance may be underestimated by an 
order of magnitude because  of the optical depths of dust continuum 
and methanol itself \citep{Hoff2018,Lee2019}. 

The difference in abundance between EC 53 and V883 Ori, despite the similar size of methanol emission, might be related to the repeated outbursts and the 2-D 
structure of the sublimation region.
The water/methanol sublimation occurs over the 2-D structure in the disk
\citep{Hoff2018,Lee2019}; the snow line extends to much larger radii at 
the disk surface than at the disk midplane. 
 During the quiescent phase, the methanol freezes back onto grain 
surfaces very efficiently in the disk midplane but much more slowly at the disk surface, where the density is much lower. 
 As a result, if there  were powerful burst events in the past in EC 53, the surface 
snow line could be very large, despite a much smaller snow line at the disk midplane because of the low current luminosity.
This will make the derived abundance of gaseous \chxoh\, low compared to the size of emission in EC 53.
In contrast, V883 Ori has an ongoing outburst with the luminosity greater than that of EC 53 in the quiescent phase by two orders of magnitude. As a result, the snow line is large even at the disk midplane, resulting in the higher abundance of gaseous methanol in V883 Ori.
 
In summary, the different timescales for freeze-out between the disk midplane and surface may explain why the methanol emission region extends far beyond the location of the water snow line expected from the current luminosity of EC~53, despite the low overall abundance compared with V883 Ori. 
The repeated accretion bursts in EC~53 may extend the water snow line, and the disk surface will retain the larger snow line for a longer time than the disk midplane.

Grain growth in the disk midplane could also contribute to increasing 
the radius of the snow line in the disk midplane.
 When the ratio of scale height to radius is 0.1, the density at the 
midplane near 0.1\arcsec\, is $\sim10^{11}$cm$^{-3}$.
 If we adopt the typical ISM grain size (0.1 $\mu$m) and the typical 
abundance of the dust grain with respect to the molecular hydrogen 
($\sim10^{-12}$), then all methanol molecules freeze out onto the 
dust grain within a few days \citep{Lee2004}.
 However, the freeze-out timescale could be longer than a few years when 
the maximum grain size is larger than 1~cm (see Section~\ref{subsec:dust_evol}), 
since the  total surface area of grains would then be reduced by a factor 
of $\sim$300, assuming that the total grain mass is conserved, and the 
minimum size of grain and the power-law index of grain size distribution 
are 0.1 $\mu$m and -3.5, respectively. 
 This freeze-out timescale is longer than the periodicity of 1.5 years
in the accretion onto EC~53 \citep{Yoo2017}.
 The \chxoh\, emission could trace the water snow line constructed
during the burst phase of the period, when the luminosity is four times
brighter \citep{Yoo2017} and the water snow line would then be two
times larger, or $\sim$0.06 \arcsec\,  \citep{Min2011,Jorgensen2015,Visser2015}.
 Although the \chxoh\, emission could extend further than 0.06\arcsec\, 
at the disk surface  because of the high temperature and low density in 
the irradiated disk \citep{Hoff2018,Lee2019}, the measured size of the 
\chxoh\, emission requires a more extreme (factor of $\sim$100) luminosity 
enhancement. 

\begin{figure*}
\includegraphics[width=1.00 \textwidth]{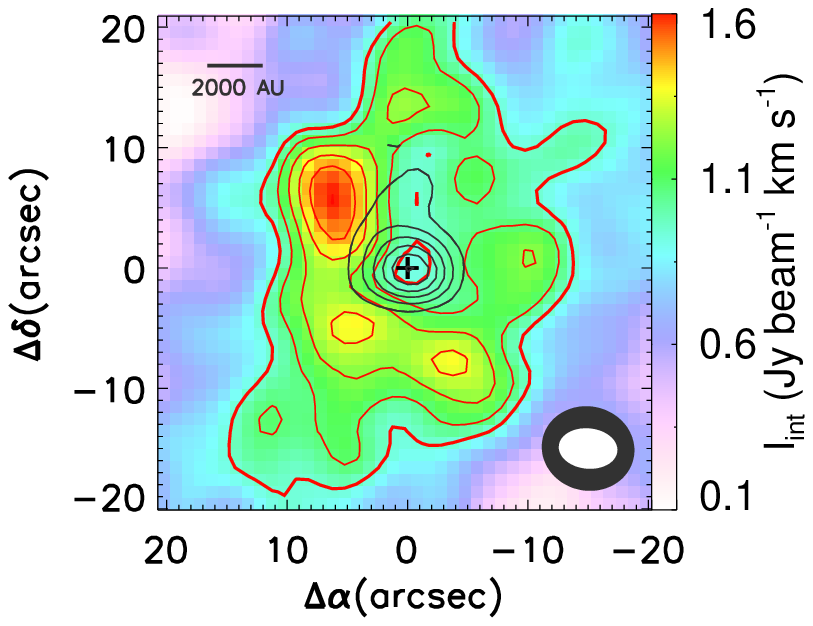}
\caption{Integrated intensity maps for the hyperfine F$_1$F=01--12 line of N$_2$H$^+$ 1--0
 (image and red contour) 
  and  the \cxo\, 3--2  (black contour) lines toward EC~53.
 The N$_2$H$^+$ image is extracted from the achieve data observed by CARMA \citep{Lee2014}
 and is integrated from -1.5 to +1.5 \kms.
The lowest level  and step of contours are 5 and 1 $\sigma$ 
(1 $\sigma$ = 0.14 Jy beam$^{-1}$ km s$^{-1}$), respectively. 
 The black contours (10$\sigma$, 20$\sigma$, 30$\sigma$, and 40$\sigma$ RMS levels, 
1$\sigma$= 132 \funitv) indicate the integrated intensity map of \cxo\, 3--2 line 
observed with ACA (W. Park et al. in prep.).
The beam sizes of CARMA and ACA are presented as the  black and white ellipses on the right
bottom corner, respectively.
The scale bar of 2000 AU is at the top left.
The N$_2$H$^+$ molecule is depleted around EC~53 where the \cxo\, line emission is detected.
}
\label{fig:classy_n2hp}
\end{figure*}

\begin{figure*}
\includegraphics[width=1.00 \textwidth]{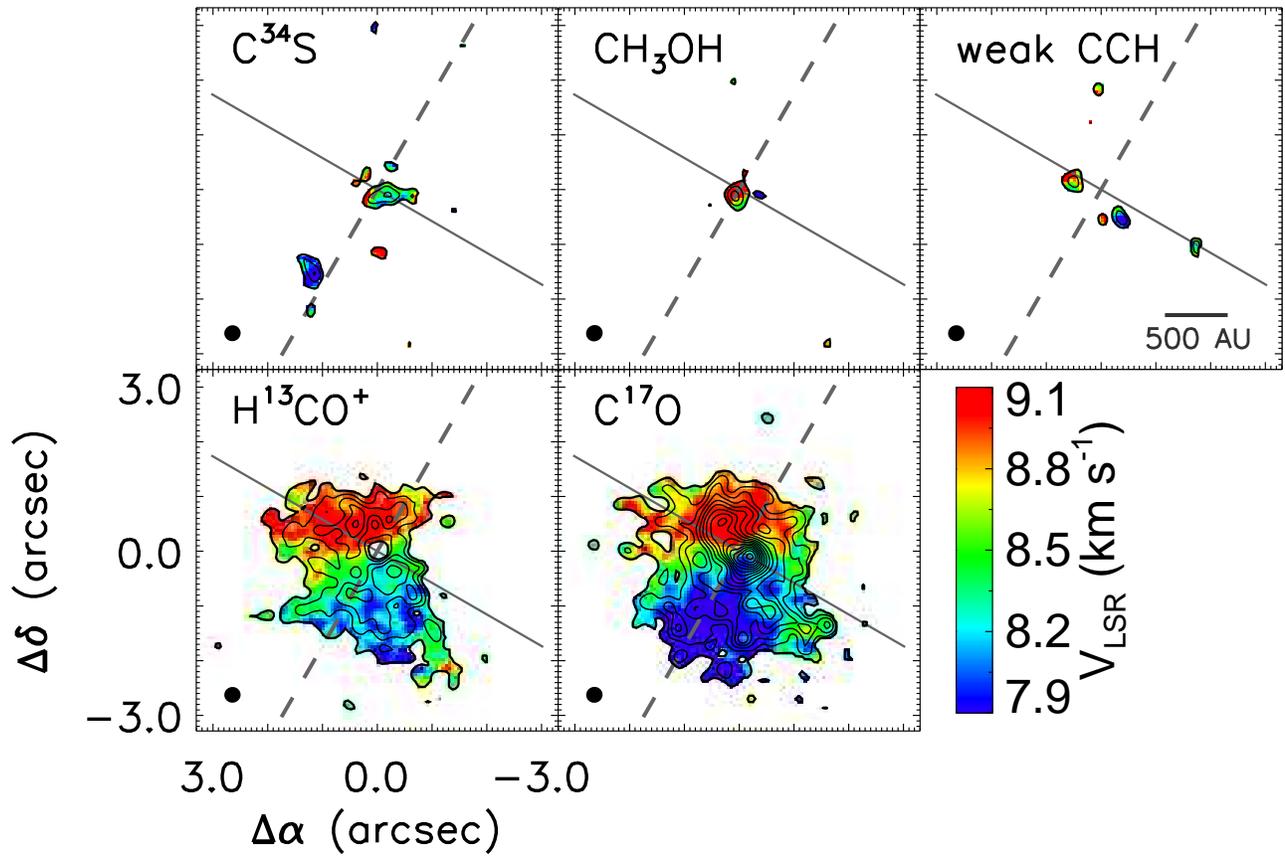}
\caption{Integrated intensity maps (contours) and intensity weighted 
velocity maps (images) for the detected molecular lines toward EC~53 
except for CCH.  The contours are the same as Figure~\ref{fig:mol_mom0}. 
}
\label{fig:mol_mom1}
\end{figure*}

\begin{figure*}
\includegraphics[width=1.00 \textwidth]{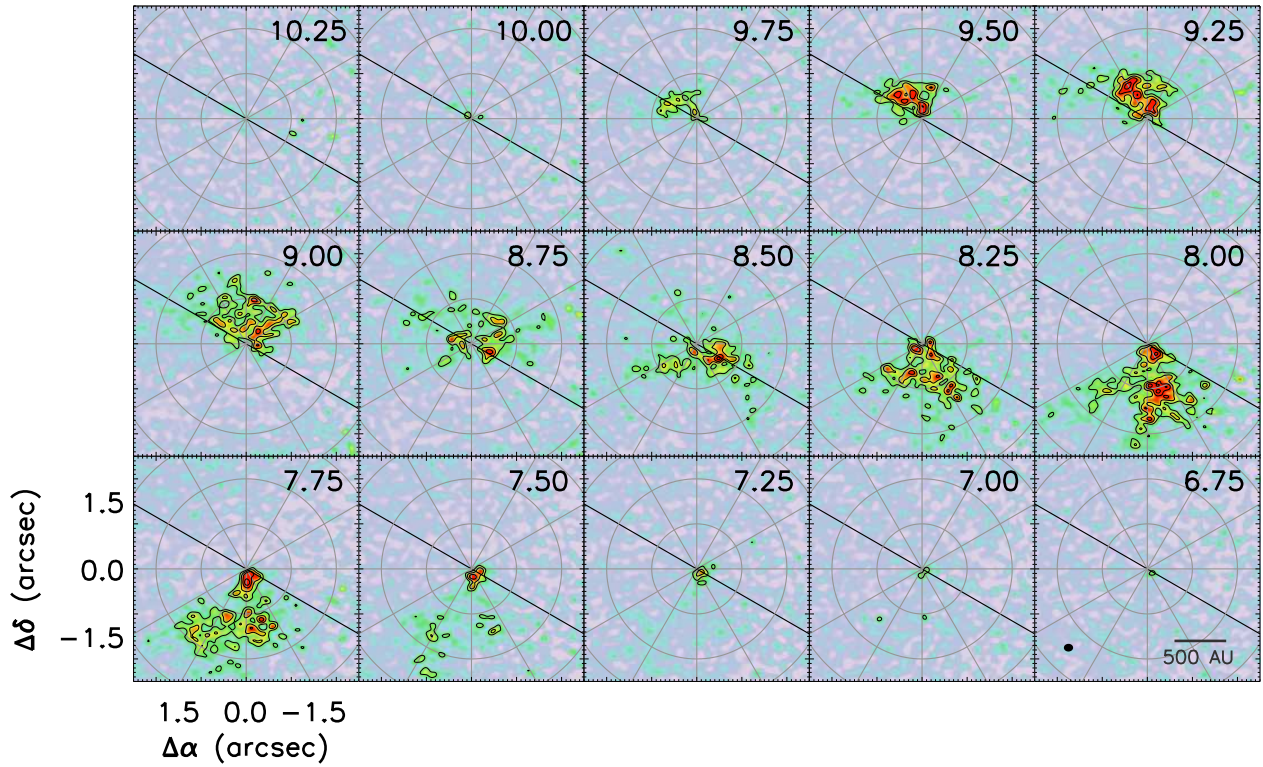}
\caption{Velocity channel maps of \cxo\, 3--2. The data are  extracted 
from the briggs weighting image. The contour levels start from 4 $\sigma$
and increase in step of 2 $\sigma$ (1$\sigma$= 3.85 \funit). 
 The gray lines and circles are plotted in step of 30\degree\, and 
1.0\arcsec, respectively, to identify the position of emission clearly.
The synthesized beam and the scale bar of 500 AU are presented at the bottom left 
 and right, respectively, in the panel with the velocity of  6.75~\kms. 
}
\label{fig:chan_c17o}
\end{figure*}

\begin{figure*}
\includegraphics[width=1.00 \textwidth]{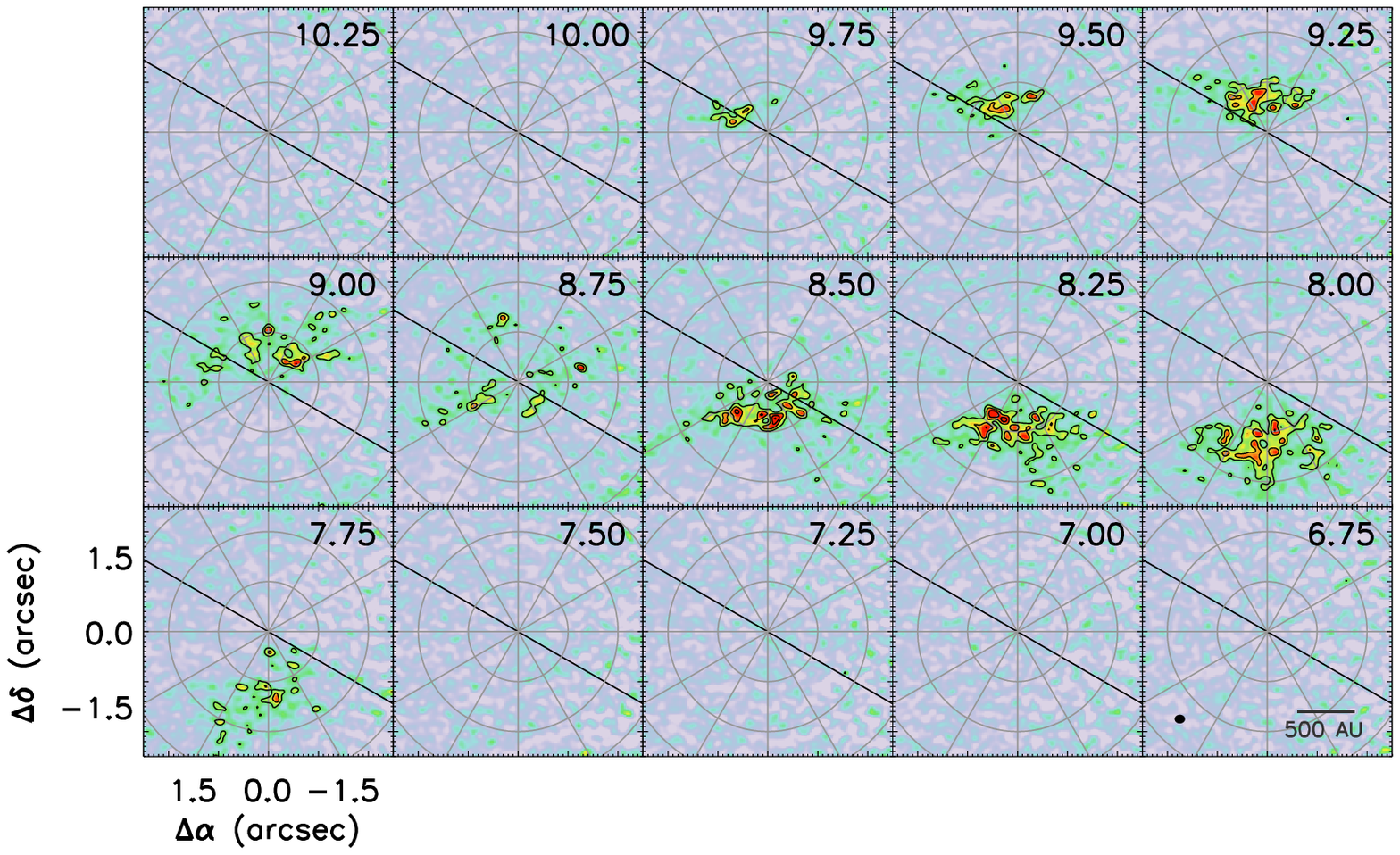}
\caption{Same as Figure~\ref{fig:chan_c17o} except for  \hcop\, 4--3. 
  The RMS noise level (1 $\sigma$) is 4.13 \funit.
}
\label{fig:chan_hcop}
\end{figure*}

\begin{figure*}
\includegraphics[width=1.00 \textwidth]{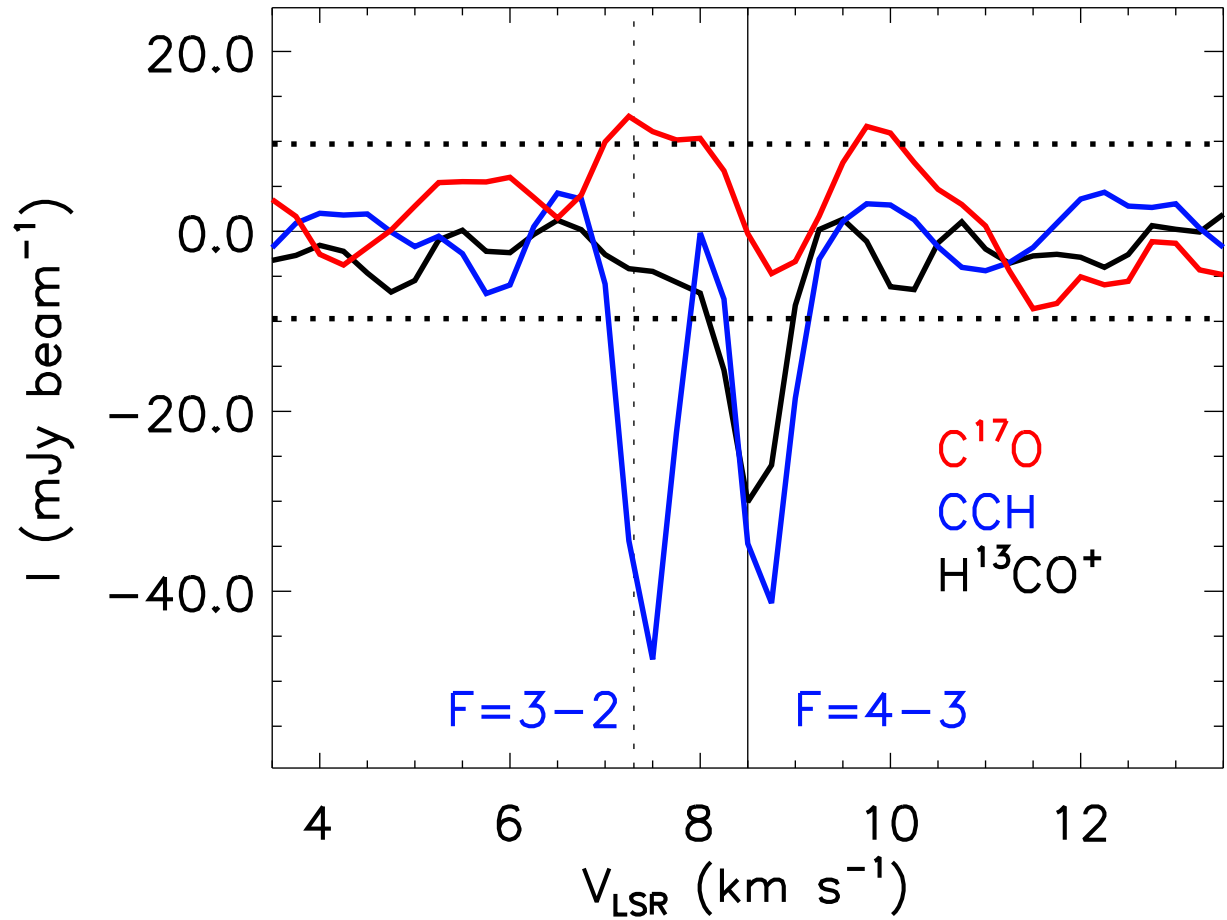}
\caption{The spectra of \hcop\, (black), CCH (blue), and \cxo\, (red) 
at the continuum peak.
 The spectra are extracted from  the briggs weighting images and Hanning-smoothed. 
 The solid vertical black line indicates the source velocity (8.5 \kms). 
 The dotted vertical line indicates the line center of the CCH F= 3--2.
 The dotted horizontal black lines show $\pm$ 3 $\sigma$ of the \cxo\, 
(1$\sigma$ = 3 \funit).
}
\label{fig:mol_spec_cen}
\end{figure*}

\begin{figure*}
\includegraphics[width=1.00 \textwidth]{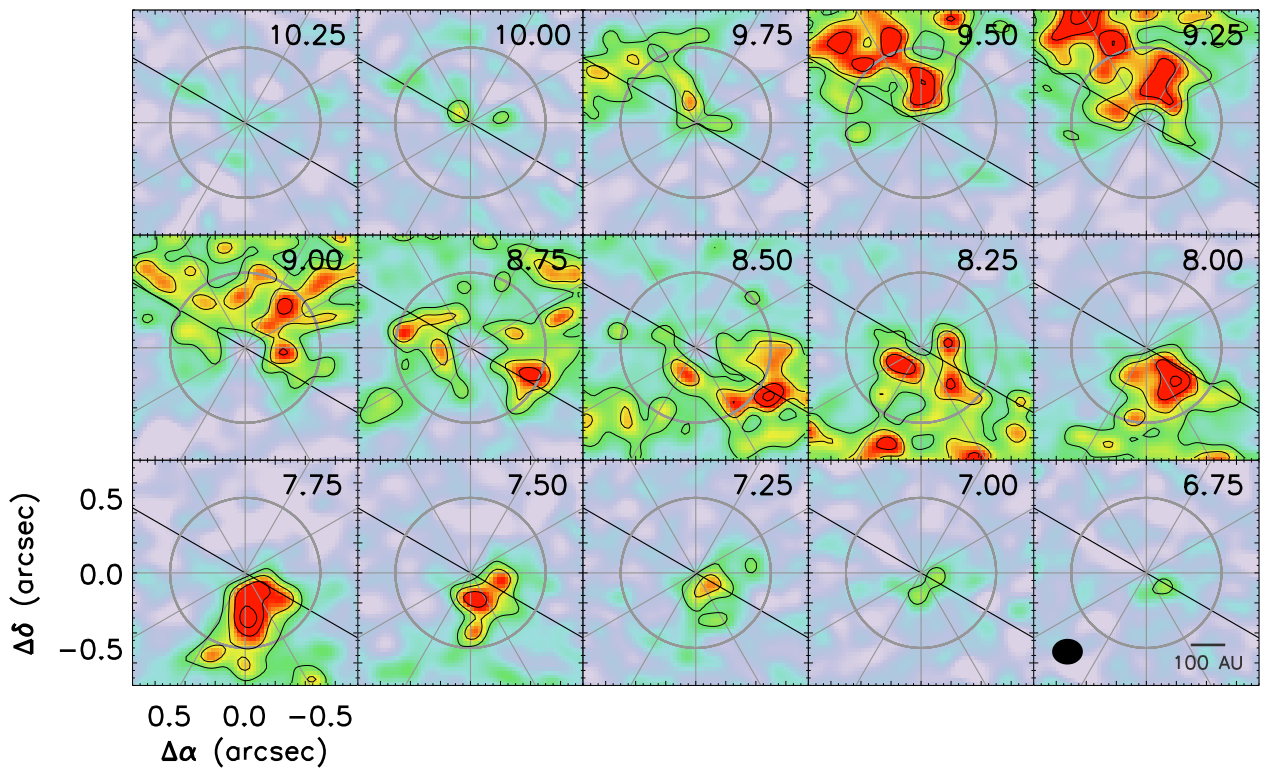}
\caption{Zoom-in of Figure~\ref{fig:chan_c17o}. The gray circle indicates 
the radius of 0.5\arcsec.
}
\label{fig:chan_c17o_zi}
\end{figure*}

\begin{figure*}
\includegraphics[width=1.00 \textwidth]{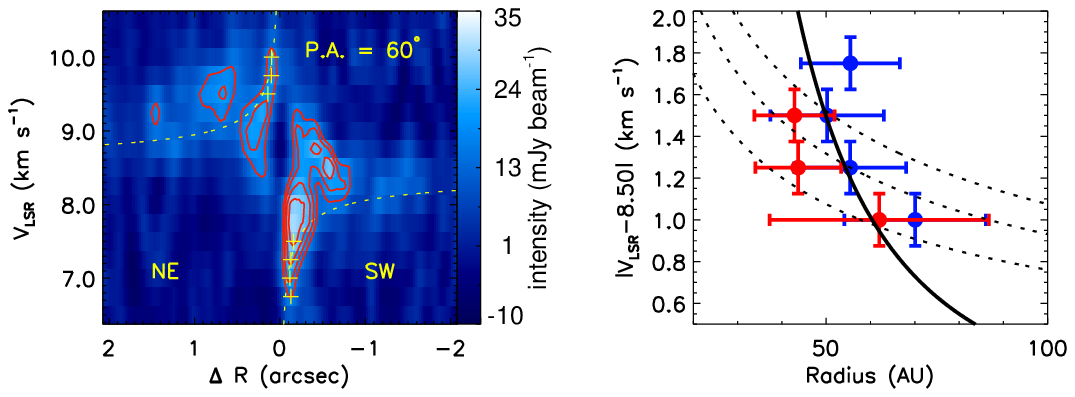}
\caption{Position-velocity diagram (left) and rotation profile (right) 
of \cxo\, 3--2 along the disk direction with a position angle 
of 60~\degree. Left: The  image and red contours indicate the 
observed \cxo\, intensity extracted from the briggs weighting image. 
 The contours start from 4 $\sigma$ in step of 1 $\sigma$ (1$ \sigma$= 3.8 \funit). 
 The dashed yellow lines represent the Keplerian motion with a protostellar 
mass of 0.3~\msun. 
 Right: The red (blue) points indicate the peak positions along given red-shifted 
 (blue-shifted) velocity channels as shown in the yellow cross in the left panel. 
 Three dotted lines represent the Keplerian motion with a protostellar mass 
of 0.4 (top), 0.3 (middle), and 0.2 \msun\, (bottom), respectively.
 When these data are fit with a power-law function, the power law index is  
-2.1 $\pm$ 2.2 (black solid line).}
\label{fig:c17o_pv}
\end{figure*}

\begin{figure*}
\includegraphics[width=1.00 \textwidth]{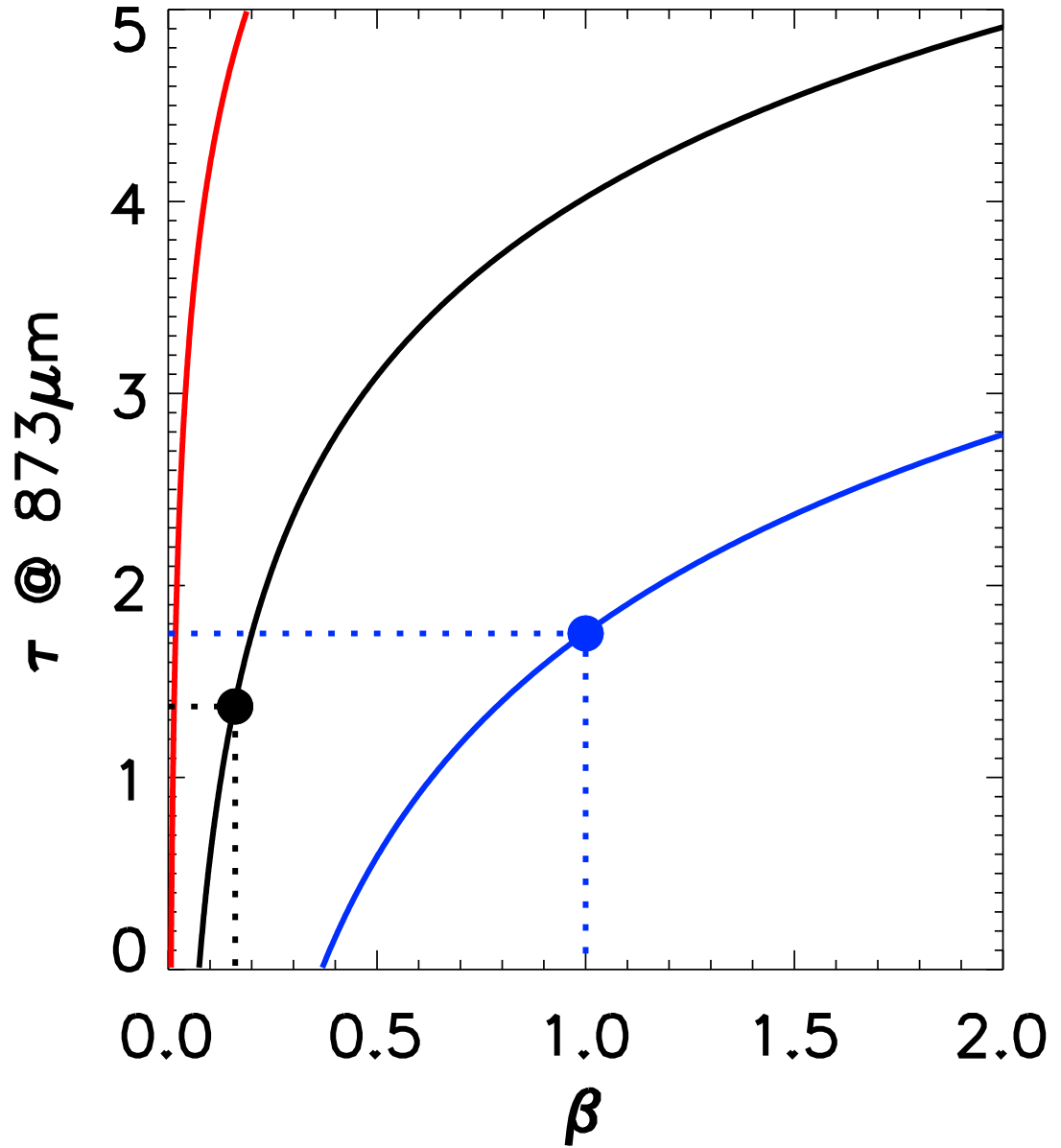}
	\caption{The optical depth at 873 $\mu$m ($\tau$) and the power-law index
	of dust opacity ($\beta$) producing the spectral index {\bf ($\alpha$)}of 1.9 for the given
dust temperatures: 20 (blue), 50 (black), and 80 K (red), respectively.
	The blue circle indicates the $\tau$ producing $\alpha$=1.9 for the  
	temperature of 20~K and $\beta$=1 used to calculate the disk mass (Section~\ref{sec:dust})
	while the black circle indicates the $\beta$ producing
	$\alpha$=1.9 for the temperature of 50~K and $\tau$=1.4} 
	derived from the thin disk model (see Table~\ref{tab:disk}). 
\label{fig:tau_beta}
\end{figure*}

\begin{figure*}
\includegraphics[width=1.00 \textwidth]{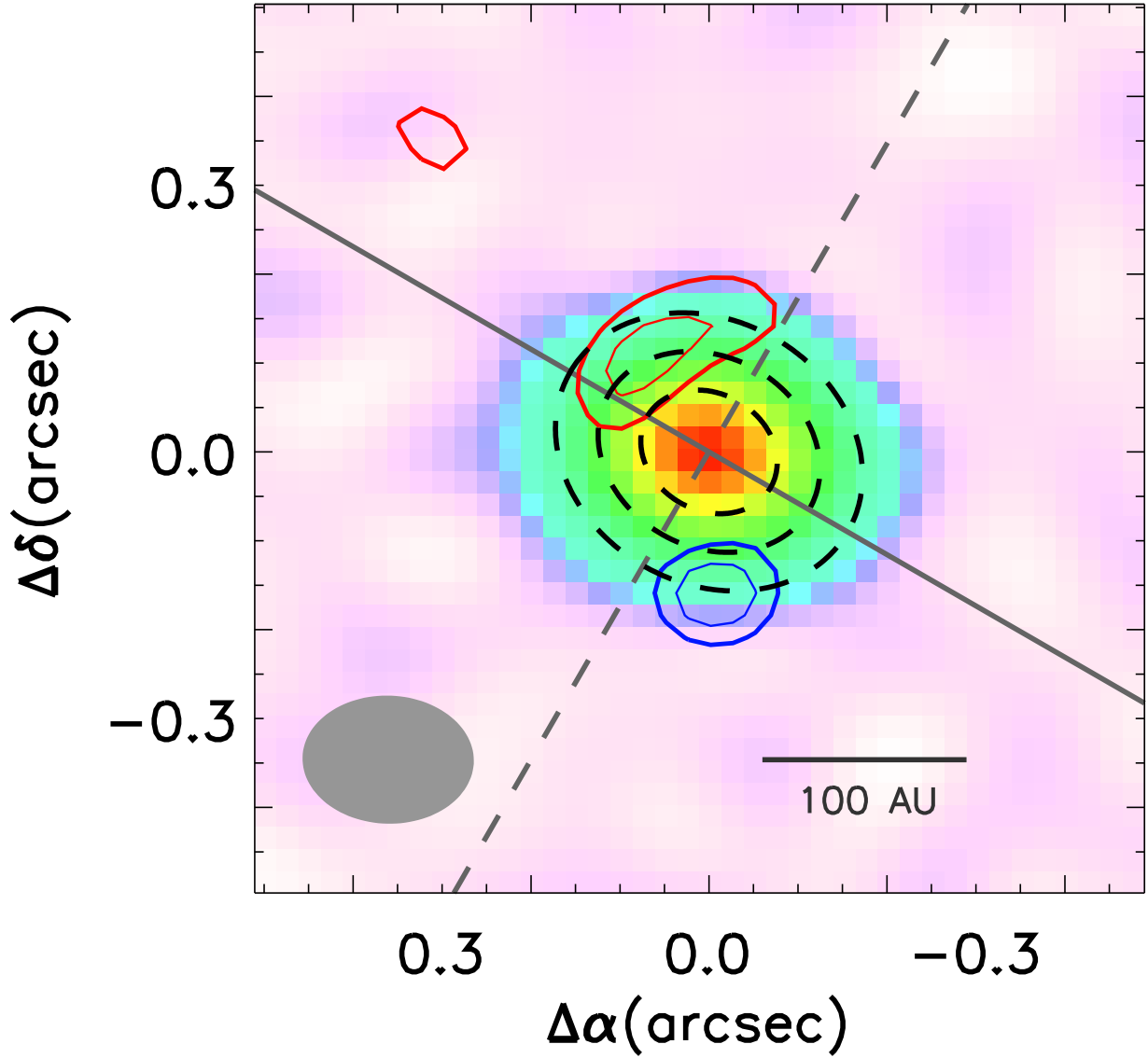}
\caption{The \cxo\, 3--2 integrated intensity map (contour) of EC~53 
produced by the uniform weighting.
 The red and blue contours indicate the red- and blue-shifted components, 
respectively, 
 integrated over the velocity from 1.0 to 1.5 \kms\, with respect to 
the source velocity.
 The lowest contour and subsequent contour step  are 5~$\sigma$ and 
1~$\sigma$, respectively  (1~$\sigma$= 3.5 \funitv).
 The background image is  the dust continuum intensity as in the left 
panel of Figure~\ref{fig:cont_uniform_alpha}.
 Three dashed ellipses correspond to the radii of 0.08, 0.13, and 0.18\arcsec, respectively. 
The beam size and 100 AU scale bar are shown in the bottom left and right corners, respectively.}

\label{fig:c17o_uniform}
\end{figure*}

\begin{figure*}
\includegraphics[width=1.00 \textwidth]{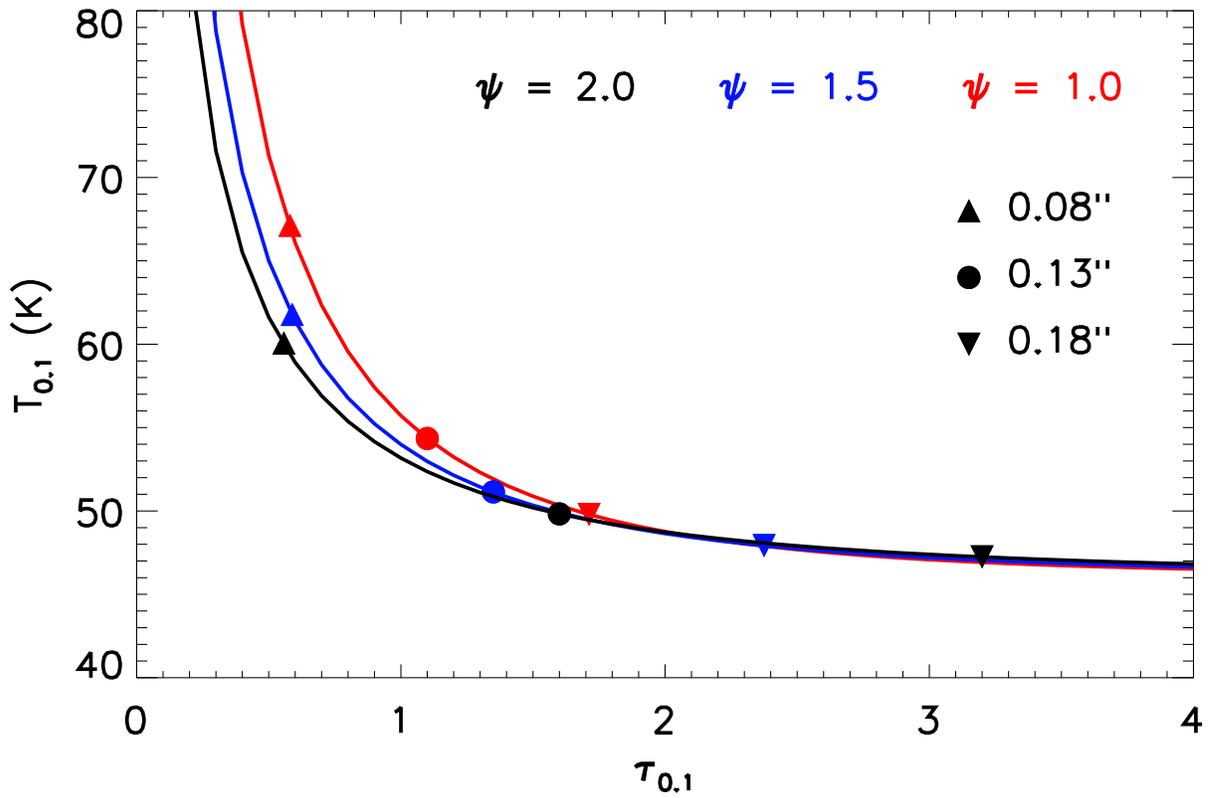}
\caption{Model parameters producing the \cxo\, emission distribution 
shown  in Figure~\ref{fig:c17o_uniform}. The lines indicate the continuum 
optical depth ($\tau_{\rm 0.1}$) and the dust temperature ($T_{\rm 0.1}$) 
at 0.1\arcsec\, which reproduce the continuum intensity (69.3 $\pm$ 0.4 \funit) 
at the continuum peak. 
Different colors indicate different power-law indices of optical depth 
(see Equation~(\ref{eq:five})). 
  The upward triangles, circles, and downward triangles indicate the model parameters
producing the \cxo\, emission peak at 0.08\arcsec, 0.13\arcsec, and 0.18\arcsec,
respectively, which are marked by the dashed ellipses in Figure~\ref{fig:c17o_uniform}.}
\label{fig:disk_model}
\end{figure*}

\begin{figure*}
\includegraphics[width=1.00 \textwidth]{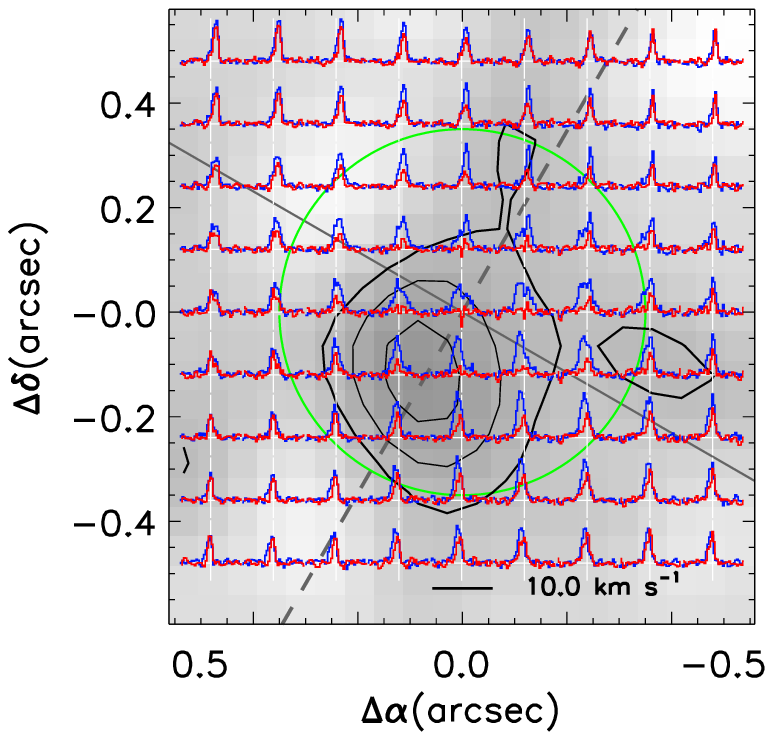}
\caption{The spectral maps of \cxo\, (blue) and \hcop\, (red) on top of 
the \chxoh\, integrated intensity map (image and contour).
 The contour levels are the same as those in Figure~\ref{fig:mol_mom1}. 
 The horizontal bar on the bottom indicates the scale of the  x-axis 
(velocity range from -5 to 5 \kms). The horizontal and vertical white 
lines indicate  the continuum level and the source velocity, respectively.
The green circle indicates the radius of 0.35\arcsec\, (153 AU at the distance of 436 pc).
}
\label{fig:spec_map}
\end{figure*}

\begin{figure*}
\includegraphics[width=1.00 \textwidth]{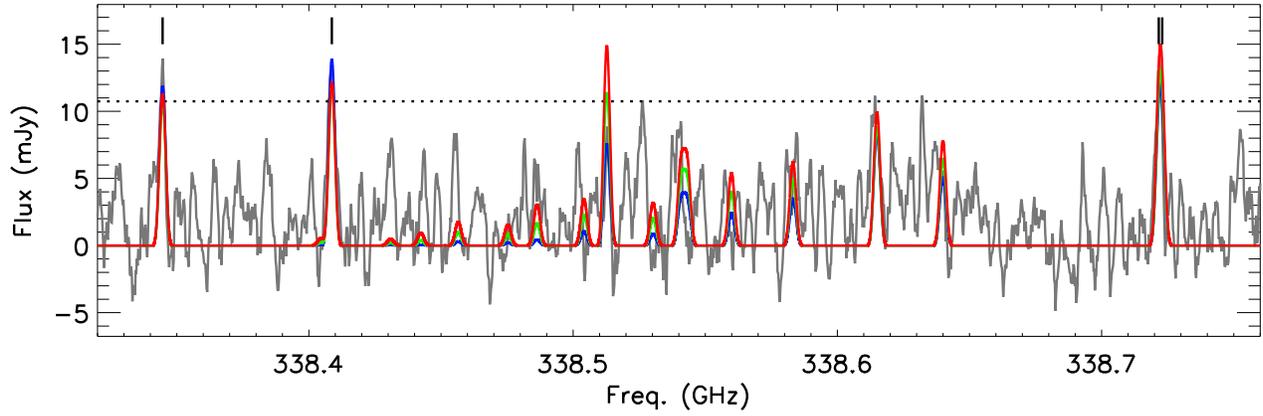}
\caption{The methanol spectra for EC~53. The gray line shows the observed 
 spectra extracted with the aperture of 0.6\arcsec.
  The spectra are smoothed with the boxcar of 1.8 \kms. The black dotted line indicates 
  the 5 $\sigma$ level (1 $\sigma$= 2.1 mJy). The four vertical black bars indicate 
  the marginally detected \chxoh\, lines presented in Table~\ref{tab:mol_lines}.
  The red, green, and blue lines show the  model spectra with the temperatures of  
  70, 50, and 30 K and with the methanol column densities of  5.2, 3.7, and 2.2 
  $\times 10^{15}$cm$^{-2}$, respectively.
}
  
\label{fig:spec_ch3oh}
\end{figure*}

 Figure~\ref{fig:classy_n2hp} shows that  the N$_2$H$^+$ line emission is 
anti-correlated with  the  \cxo\, 3--2 line emission. 
 It is hard to determine accurately the size of the N$_2$H$^+$ emission 
hole because of the low spatial resolution ($\sim$7\arcsec) of CARMA.
 However, our \cxo\, ALMA data (Figures~\ref{fig:mol_mom0} and \ref{fig:chan_c17o}) 
indicate  that the \cxo\, line  extends to $\sim$2\arcsec\,  along the 
disk midplane direction.
 In addition, in the ACA  observation (Figure~\ref{fig:classy_n2hp}), the 
full width at half maximum of the \cxo\, 3--2 emission is about 2.7\arcsec. 
 This observed CO snow line ($\sim$1000 AU) is also larger than that 
expected from the luminosities both for the quiescent and burst phases 
($\sim$250 AU and $\sim$500 AU, respectively).  
 These CO sublimation radii (25 K) are derived from the spectral energy 
distribution modelings of EC~53 (G. Baek et al. submitted.). 
 Together, the estimated sizes of the \cxo\,  emission and the  N$_2$H$^+$ 
emission hole support suggestion of repeated and/or more extreme bust 
events in the past.
  
\section{Summary}
We carried out high resolution ALMA observations toward EC~53. The main 
results are:

1) The continuum image at 873 $\mu$m shows that EC~53 has a massive disk 
(0.075 $\pm$ 0.003 M$_\odot$) with  centimeter-size grains, implying  that  
a planet could form even in the early evolutionary stage to induce the  
observed periodic variation of accretion rate.
Furthermore, the optically thick dust continuum emission in the sub-mm 
reduces the molecular line emission near the continuum peak.

(2) The high velocity components of \cxo\, 3--2 show a rotation feature 
along the disk direction, however, the current observation cannot constrain 
whether the rotation profile is associated with the Keplerian disk and/or 
an infalling envelope.
 If the Keplerian motion is adopted, the rotation profile
indicates that EC~53 has a protostellar mass of $\sim$0.3 $\pm$ 0.1 ~M$_\odot$.

(3) The anti-correlation of \chxoh\, and \hcop\, near the continuum peak 
could be associated with the destruction of \hcop\, by  evaporated water 
molecules. 

(4) The observed methanol and CO emissions are more extended than the snow 
lines expected due to the current luminosity, implying repeated and/or 
more extreme accretion burst events in the past.  The luminosity and therefore the abundance of \chxoh\ is two orders of magnitude lower than measured from disks undergoing larger outbursts, perhaps because the \chxoh\ is frozen out at the disk midplane but is still in the gas phase in the lower-density disk surface.

\acknowledgments
This work was motivated and performed by the JCMT-Transient Team as part of a comprehensive effort to obtain and interpret time-domain sub-mm observations of star-forming regions. The authors appreciate the important contribution of the JCMT Transient Team members in observing, calibrating, and making available the JCMT sub-mm observations used for this paper. This work was supported by the Basic Science Research Program through the
 National Research Foundation of Korea (grant No. NRF-2018R1A2B6003423) and 
the Korea Astronomy and Space Science Institute under the R\&D program 
supervised by the Ministry of Science, ICT and Future Planning.
S. L. was also supported by NAOJ ALMA Scientific Research Grant Number of 2018-10B.
  Y.A. was supported by KAKENHI Grant Number 18H05222 and NAOJ ALMA Scientific Research Grant Number of 2019-13B.  GJH is supported by general grant 11773002 awarded by the National Science Foundation
of China.
D.J. is supported by the National Research Council of Canada and by an NSERC 
Discovery Grant.
%
%

\bibliography{ec53}
\bibliographystyle{aasjournal}
\clearpage

\begin{deluxetable}{lccccc}
\tablewidth{0.7 \textwidth}
\centering
\tablecaption{Detected molecular lines \label{tab:mol_lines}}
\tablehead{\colhead{Molecule} &
       \colhead{Frequency (GHz)} &
       \colhead{Transition} &
       \colhead{log$_{10}$ A$^{a}$ (s$^{-1}$)} &
      \colhead{E$_{\rm up}^{b}$ (K) }  &
       \colhead{g$_{\rm up}^{c}$} } 
\startdata
C $^{17}$O     & 337.060988$^{d}$   &  J= 3--2                & -5.63433 & 32.4 & 12 \\
C$^{34}$S     & 337.396459          & J= 7--6                & -3.11804 & 50.2 & 15 \\
CH$_3$OH      & 338.3446280$^{e}$ & 7(-1,7)--6(-1,6)        & -3.77817 & 70.5 & 15 \\
              & 338.4086810$^{e}$ & 7(0,7)--6(0,6)++        & -3.76902 & 65.0 & 15 \\
             & 338.7216300$^{e}$ & 7(2,6)--6(2,5)        & -3.80441 & 90.9 & 15 \\
              & 338.7229400$^{e}$ & 7(-2,6)--6(-2,5)        & -3.80441 & 90.9 & 15 \\
H$^{13}$CO$^+$ & 346.9983440        &  J= 4--3                & -2.48309 & 41.6 & 9 \\
CCH            & 349.3992756$^{f}$ & N= 4--3, J= 7/2--5/2, F= 4--3 & -3.90313 & 41.9 & 9 \\ 
              & 349.4006712$^{f}$ & N= 4--3, J= 7/2--5/2, F= 3--2 & -3.92139 & 41.9 & 7 \\ 
              & 349.4146425$^{e,g}$ & N= 4--3, J= 7/2--5/2, F= 3--3 & -5.15397 & 41.9 & 7 \\ 
              & 349.6036139$^{e,g}$ & N= 4--3, J= 7/2--7/2, F= 4--4 & -5.28599 & 41.9 & 9  
\enddata
\tablecomments{The molecular information is adopted from CDMS 
  database \citep{CDMS2001}.}
\tablenotetext{a}{Einstein A-coefficient.}
\tablenotetext{b}{Energy state of the upper level.}
\tablenotetext{c}{Statistical weight of the upper level. }
\tablenotetext{d}{Frequency of the strongest hyper-fine line in \cxo\, 3--2.}
\tablenotetext{e}{Marginally detected line.}
\tablenotetext{f}{Strong CCH line.}
\tablenotetext{g}{Weak CCH line.}
\end{deluxetable}


\begin{deluxetable}{lcccc}
\tablewidth{1. \textwidth}
\centering
\tablecaption{Thin disk model for the Continuum emission \label{tab:disk} }
\tablehead{\colhead{$\psi^{a}$} &
       \colhead{    } &
       \colhead{\quad\quad\quad 1.0\quad\quad\quad} &
       \colhead{\quad\quad\quad 1.5\quad\quad\quad} &
       \colhead{\quad\quad\quad 2.0\quad\quad\quad} }
\startdata
  $T_{\rm 0.1}^{a}$ & (K)  & 52 & 50 & 49 \\
  $\tau_{\rm 0.1}^{a}$ &   & 1.1  & 1.4  & 1.6 \\
  $r_{\rm max}^{b}$ & (\arcsec) & 0.14 &  0.13 &  0.13 \\
  $T_{\rm med}^{c}$ & (K) & 53 &  51 &  50 \\
  $M_{\rm disk}^{d}$ & (\msun) & 0.05 & 0.10 & 0.23
\enddata
\tablenotetext{a}{Best fit parameters in the thin disk models 
  (see Figure~\ref{fig:disk_model})}
\tablenotetext{b}{The truncated radius of disk producing the 
  observed continuum flux (139.7 mJy).}
\tablenotetext{c}{The surface area weighted temperature within a disk ( $r < r_{\rm max}$).}
\tablenotetext{d}{The disk mass from the thin disk model with the 
  dust opacity of 3.4~cm$^2$~g$^{-1}$ \citep{Beckwith1990} and the gas to dust ratio of 100. }                                 
\end{deluxetable}

\clearpage

\clearpage

\clearpage


\clearpage

\clearpage

\clearpage

\clearpage

\clearpage

\clearpage
\clearpage

\clearpage

\clearpage

\clearpage

\clearpage
\clearpage

\end{document}